\documentclass[aps,pre,reprint,10pt,twocolumn,showpacs,nofootinbib]{revtex4-1} %aps,showpacs,showkeys,groupedaddress,superscriptaddress

\usepackage{amsmath}
\usepackage{amsfonts}
\usepackage{amssymb}
\usepackage{graphicx}
\usepackage{subfigure}
\usepackage[pdfborder={0 0 0}]{hyperref}  
\usepackage{color}
\usepackage[latin1]{inputenc}
\usepackage{float}
\usepackage{comment}

\begin{document}

\title{Entanglement of Spin-$1/2$ Heisenberg Antiferromagnetic Quantum Spin Chains}
 
\begin{abstract}
The quantum entanglement measure is determined, for the first time, for a collection 
of spin-$1/2$ arranged in a infinite chain with finite temperature and 
applied to a single-crystal $\beta-\mathrm{T_eVO_4}$. The physical 
quantity proposed here to measure the entanglement is the distance between states by 
adopting the Hilbert-Schmidt norm. We relate the distance between states with the
magnetic susceptibility. The decoherence temperature, above which the entanglement is suppressed, 
is determined for a system. A correlation among their decoherence temperatures and 
their respective exchange coupling constants is established; moreover, it is conjectured 
that the exchange coupling protects the system from decoherence as temperature increases.
\end{abstract}

\author{Saulo Luis Lima da Silva}

\email{saulo.silva@ufv.br}

\affiliation{Universidade Federal de Vi\c cosa,\\
Departamento de F\'\i sica -- Campus Universit\'ario\\
Avenida Peter Henry Rolfs s/n -- 36570-900 --
Vi\c cosa -- MG -- Brazil.}

\maketitle

%%%%%%%%%%%%%%%%%%%%%%%%%%%%%%%%%%%%%
\section{Introduction}
%%%%%%%%%%%%%%%%%%%%%%%%%%%%%%%%%%%%%
Currently studying entanglement in condensed matter systems is of great interest.
This interest stems from the fact that some behaviors of such systems can only be 
explained with the aid of entanglement. The magnetic susceptibility at low 
temperatures, quantum phase transitions, chemical reactions are examples where 
the entanglement is key ingredient for a complete understanding of the system.
Furthermore, in order to produce a quantum processor, the entanglement of study 
condensed matter systems becomes essential. In condensed matter, said magnetic 
materials are of particular interest. Among these we will study the 
antiferromagnetism which are described by Heisenberg model.

We use the Hilbert-Schmidt norm for measuring the distance between quantum states.
The choice of this norm was due mainly to its application simplicity and 
strong geometric appeal. The question 
of whether this norm satisfies the conditions desirable for a good measure of 
entanglement was discussed in 1999 by C. Witte and M. Trucks \cite{ref1}. They 
showed that the norm of Hilbert-Schmidt is not increasing under completely positive 
trace-preserving maps making use of the Lindblad theorem \cite{ref2}. M. Ozawa 
argued that this norm does not satisfy this condition by using an exemple
of a completely positive map which can enlarge the Hilbert-Schmidt norm between two states 
\cite{ref3}. However this does not prove the fact that the entanglement measure
based on the Hilbert-Schmidt norm is not entangled monotone. This problem has 
come up in several contexts in recent years. Superselection structure of dynamical semigroups 
\cite{ref3.1}, entropy production of a quantum chanel \cite{ref3.2}, condensed matter theory 
\cite{ref3.3} and quantum information \cite{ref3.4, ref15, ref10} are some examples. Several 
authors have been devoted to this issue in recent years \cite{ref3.5, ref3.6, ref3.7, ref3.8, ref3.9}
and other work on this matter is in progress by the author and collaborators. 

The study of entanglement in Heisenberg chains is of great interest in physics and has 
been done for several years. In the early 2000s, K. M. O'Connor, W. K. Wootters, X. Wang and 
P. Zanardi showed how to get the density matrix of a chain of Heisenberg whose Hamiltonian 
commutes with the z component of the total spin \cite{ref4, ref5}. M. Wie\'sniak, V. Vedral 
and C. Bruckner showed in 2005 how to relate entanglement with the magnetic susceptibility 
\cite{ref6}. In 2008 S. M. Aldashin studied the entanglement in dimer systems \cite{ref7}.
The authors made use of Blaney and Bowers equation \cite{ref8} to relate the entanglement 
with the magnetic susceptibility of the system. The entanglement of a dimer-trimer system was 
studied experimentally using magnetic susceptibility in 2008 by M. Souza \textit{et al.} 
\cite{ref9}. Del Cima \textit{et al.} had in 2015 entanglement for the trimer compound, relating it 
to the magnetic susceptibility of the material. Also obtained the entanglement 
depending on the temperature and the critical temperature of entanglement for two compounds 
that have not yet been studied in this respect \cite{ref10}.

In the present work, we have studied a Heisenberg infinite chain system from the perspective of entanglement. We can find a discussion of infinite chain in \cite{ref4, ref6, ref11, ref23}. However, these discussions 
are limited to zero temperature or are not accurate results\footnote{For use Entanglement Witnesses (EW)}. In this paper, we have, accurately, detected entangled states in finite temperature. For this purpose, we make use of Bonner-Fisher model \cite{ref12}, and work of Eggert \textit{et l.} 
\cite{ref13}, to relate the entanglement with the magnetic susceptibility of the system. In addition, we have, for the first time, an entanglement in the compound $\beta-\mathrm{T_eVO_4}$. Present entanglement as a function of temperature
as well as the critical temperature of entanglement.

The paper is structured as follows. In Section \ref{sec:Calcu} we present 
the method we will use to quantify entanglement. In Section \ref{sec:entangled chain} 
we calculate the entanglement for a set of $N$ particles arranged in a ring, 
after we will generalize this for a infinite chain. This result was applied to
the $\beta - \mathrm{T_eVO_4}$ compound. Finally, Section  \ref{sec:Conc} is 
intended to conclusions.

%%%%%%%%%%%%%%%%%%%%%%%%%%%%%%%%%%%%%
\section{Calculating the entanglement}
\label{sec:Calcu}
%%%%%%%%%%%%%%%%%%%%%%%%%%%%%%%%%%%%%
Following \cite{ref10}, we consider a system that satisfies 
the Heisenberg Hamiltonian
\begin{equation}\label{eq0}
H = -J \sum_i \boldsymbol{S}_i \cdot \boldsymbol{S}_{i+1}, 
\end{equation}
where $J$ is the exchange constant and $\boldsymbol{S}_i$ is the spin ope\-ra\-tor of the 
site $i$. The Hamiltonian commutes with the $z$-component of total 
spin $[H,S^z]=0$, there is no coherent superposition of states and $J<0$. This allows us to write the reduced density matrix \cite{ref4}
\begin{equation}\label{state}
\rho_{i,i+1}=
\begin{pmatrix}
v & 0 & 0 & 0 \\
0 & w & z & 0 \\
0 & z^* & w & 0 \\
0 & 0 & 0 & v
\end{pmatrix}.
\end{equation}
where the indices $i$ and $i+1$ refer to the site $i$ and its nearest 
neighbor,  respectively. We can relate the elements of the reduced density matrix with 
the correlation function per site as follows \cite{ref5}:
\begin{equation}\label{eq3}
v=\frac{1}{4}+\langle S_i^zS_{i+1}^z\rangle,
\end{equation}
and
\begin{equation}
z=\langle S_i^xS_{i+1}^x \rangle + \langle S_i^yS_{i+1}^y \rangle + 
i\langle S_i^xS_{i+1}^y \rangle - i\langle S_i^yS_{i+1}^x \rangle.
\end{equation}
In the absence of an external magnetic field the system is isotropic; thus
\begin{equation}\label{eq4}
z=2\langle S_iS_{i+1} \rangle.
\end{equation}
The system has the dimension of the Hilbert space associated 
$2\otimes2$. Therefore, we can use the criterion of 
Peres-Horodecki \cite{ref14}. The eigenvalues of the 
partial transpose of $\rho_{i,i+1}$ are
\begin{equation}
\{w,w,v-\vert z\vert,v+\vert z \vert \}.
\end{equation}
Hence, we can see that the system is entangled when
\begin{equation}\label{eq1}
v < \vert z \vert,
\end{equation}
otherwise, the system is separable.

\begin{figure}
 \centering
 \includegraphics[width=6.5 cm]{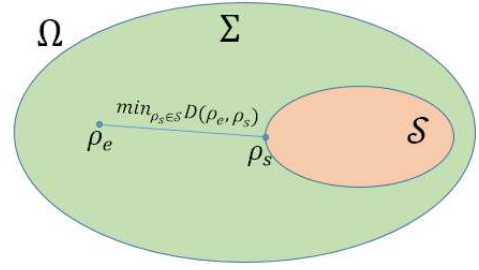}
 \caption{Distance between states as a measure of entanglement.}
 \label{f:0}
\end{figure}

Knowing what are the separate and entangled states, 
we can calculate how much each state is entangled. 
We use the Hilbert-Schmidt norm \cite{ref1, ref15} to determine the 
distance between states in order to evaluate the degree of 
entanglement of the system. Let us consider
a set $\Omega$ of all density matrices. It consists of two 
disjunctive subsets: the subset of separable states $\mathcal{S}$ 
and the subset of entangled states $\Sigma=\Omega - \mathcal{S}$.
In this case, the entanglement is given by
\begin{align}
\mathcal{E}(\rho_{\rm e})=\min_{\rho_{\rm s} \in \mathcal{S}}\,D(\rho_{\rm s}, \rho_{\rm e}),
\end{align}
where $D$ is the distance between the density matrices $\rho_{\rm e}$ (entangled)
and the set of separable states $\mathcal{S}$ (see Figure \ref{f:0}). 
In other words, we take the degree of entanglement as the shortest distance between a given 
entangled state and the set of reduced density matrices separable. 
$\rho_{\rm s}$ and $\rho_{\rm e}$ will be given by
\begin{equation}
\rho_{\rm s}=
\begin{pmatrix}
v_{\rm s} & 0 & 0 & 0 \\
0 & w_{\rm s} & z & 0 \\
0 & z^* & w_{\rm s} & 0 \\
0 & 0 & 0 & v_{\rm s}
\end{pmatrix}, \,\,\,\, \mathrm{and} \,\,\,\, \rho_{\rm e}=
\begin{pmatrix}
v_{\rm e} & 0 & 0 & 0 \\
0 & w_{\rm e} & z & 0 \\
0 & z^* & w_{\rm e} & 0 \\
0 & 0 & 0 & v_{\rm e}
\end{pmatrix}.
\end{equation}
Hence, the entanglement is
\begin{equation}\label{eq2}
\mathcal{E}(\rho_{\rm e})={\min} \sqrt{Tr[(\rho_{\rm s} - \rho_{\rm e})^2]}=2\,{\min}\,\vert v_{\rm s} - v_{\rm e} \vert.
\end{equation}
It is easy to see that (\ref{eq2}) is minimal when $v_{\rm s}=\vert z \vert$.
Thus, the entanglement of the system is given by
\begin{equation}\label{eq5}
\mathcal{E}(\rho_{\rm e})={\max}\left[0,2(\vert z \vert - v_{\rm e})\right].
\end{equation}
The term ``${\max}$'' was introduced to avoid negative 
entanglement. Replacing (\ref{eq3}) and (\ref{eq4}) in (\ref{eq5}) we obtain
\begin{equation}\label{eq6}
\mathcal{E}(\rho_{\rm e})=2\,{\max}\left[0,\left(2\vert \langle S_iS_{i+1}\rangle \vert 
- \frac{1}{4}-\langle S_iS_{i+1}\rangle\right) \right].
\end{equation} 
Since the Hamiltonian $H$ commutes with the spin component along the $z$
direction, $S^z$, one can show that the magnetic susceptibility
along a given direction $\alpha$ can be written as \cite{ref6}
\begin{equation}\label{eq7}
\chi^\alpha(T)=\frac{(g\mu_B)^2}{k_B T}\left(\sum_{j,k=1}^{N}\langle S_j^\alpha S_k^\alpha\rangle 
- \left\langle \sum_{k=1}^{N}S_k^\alpha \right\rangle^2 \right),
\end{equation}
and
\begin{equation}\label{eq8}
\overline{\chi}(T)=\frac{\chi^x +\chi^y +\chi^z}{3}.
\end{equation}
It should be noticed that for optical lattices, the variance in 
(\ref{eq7}) can be directly measured without to stand in need of the 
magnetic susceptibility \cite{ref16}.

If $N$ is even, we have
\begin{equation}\label{eq13}
\langle S_i S_j \rangle_{\rm even}=\frac{12k_BT\overline{\chi}-3N(g\mu_B)^2}{8(N-1)(g\mu_B)^2},
\end{equation}
for $N$ odd
\begin{equation}\label{eq14}
\langle S_i S_j \rangle_{\rm odd}=\frac{12k_BT\overline{\chi}-(3N-1)(g\mu_B)^2}{8(N-1)(g\mu_B)^2},
\end{equation}
or
\begin{equation}
\langle S_i S_j \rangle_{\rm odd}=\langle S_i S_j \rangle_{\rm even}+\frac{1}{8(N-1)}~.
\end{equation}

Using (\ref{eq13}), (\ref{eq14}) and (\ref{eq6}) we can relate 
the entanglement of the system of interest with the magnetic 
susceptibility of system. Note that $\langle S_i S_j \rangle_{\rm odd}=\langle S_i S_j \rangle_{\rm even}$ 
when $N \rightarrow \infty$. 

%%%%%%%%%%%%%%%%%%%%%%%%%%%%%%%%%%%%%
\section{Quantum Correlations}
\label{sec:entangled chain}
%%%%%%%%%%%%%%%%%%%%%%%%%%%%%%%%%%%%%
Replacing (\ref{eq13}) in (\ref{eq6}) we have the entanglement for a ring with $N$ even spins
\begin{align}\label{eq15}
\mathcal{E}(\overline{\chi})&=2\,{\max}\left[0,\left(2\left\vert\frac{12k_BT\overline{\chi}
-3N(g\mu_B)^2}{8(N-1)(g\mu_B)^2}\right\vert
-\frac{1}{4} \right.\right. \nonumber \\[3mm]
&~~\left.\left. -\frac{12k_BT\overline{\chi}-3N(g\mu_B)^2}{8(N-1)(g\mu_B)^2}\right)\right].
\end{align}
Similarly, replacing (\ref{eq14}) in (\ref{eq6}) we have the entanglement for a ring with $N$ odd spins
\begin{align}\label{eq16}
\mathcal{E}(\overline{\chi})&=2\,{\max}\left[0,\left(2\left\vert\frac{12k_BT\overline{\chi}
-(3N-1)(g\mu_B)^2}{8(N-1)(g\mu_B)^2}\right\vert
-\frac{1}{4} \right.\right. \nonumber \\[3mm]
&~~\left.\left. -\frac{12k_BT \overline{\chi}-(3N-1)(g\mu_B)^2}{8(N-1)(g\mu_B)^2}\right)\right].
\end{align}
In the limit $N \rightarrow \infty$ the Eqs. (\ref{eq15}) and (\ref{eq16}) become equal.

In Ref.\cite{ref12}, Bonner and Fisher have shown that magnetic susceptibility for a infinite chain is
\begin{align}\label{eq17}
\chi(T)\!=\!\frac{N(g\mu_B)^2\bigl(0.25+0.074795x+0.075235x^2\bigr)}
{k_BT\bigl(1.0+0.9931x+0.172135x^2+0.757825x^3\bigr)}, \nonumber \\
\end{align}
where $x=\vert J \vert / k_BT$. In the low temperature regime Eggert \textit{et al.} 
\cite{ref13} show that an asymptotic dependence of the magnetic susceptibility 
with $(\mathrm{ln}\,T)^{-1}$. At high temperatures both results agree.

Taking the limit $T \rightarrow 0$ and replacing (\ref{eq17}) in (\ref{eq16}), we obtain
the entanglement of an infinite chain at $T=0$.
In this case, using the susceptibility 
of Eggert \textit{et al.} or of Bonner-Fisher give the same result
\begin{equation}\label{eq18}
\mathcal{E}(T=0)=\frac{7}{16}.
\end{equation}
Note that this result is independent of $J$. In other words, it is independent 
of the material. This result is significantly close to that found by O' Connor 
\textit{et al.} in Ref.\cite{ref4}, where the authors tried to answer the question about
what extent each pair of nearest neighbors can be entangled. But we can not say 
that the results found by them is optimal. 

If we consider $T>0$, at low temperature we should use the result of the Ref.\cite{ref13};
but most of the materials in this temperature range undergoes phase transitions that prevent 
the use of the one-dimensional model. Fortunately, 
this temperature range is small ($T<5\,\mathrm{K}$)! At high temperatures we can use the result 
of Bonner and Fisher for thermal entanglement.

%%%%%%%%%%%%%%%%%%%%%%%%%%%%%%%%%%%
\subsection{$\beta - \mathrm{T_eVO_4}$ compound}
\begin{figure}
 \centering
 \includegraphics[width=9.4 cm]{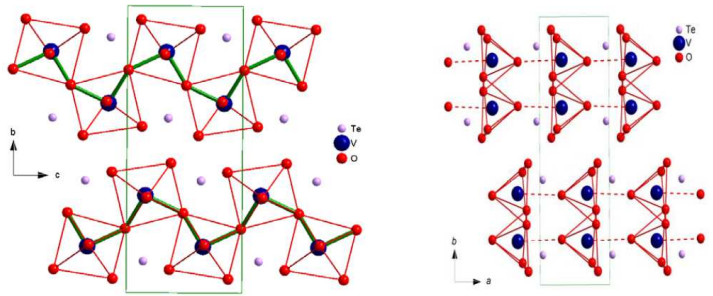}
 \caption{(Color online) Crystal structure of $\beta - \mathrm{T_eV)_4}$, viewed along \textit{a} 
 (left) and \textit{c} (right) axis \cite{ref17}.}
 \label{f:1}
\end{figure}

\begin{figure}
 \centering
 \includegraphics[width=6 cm]{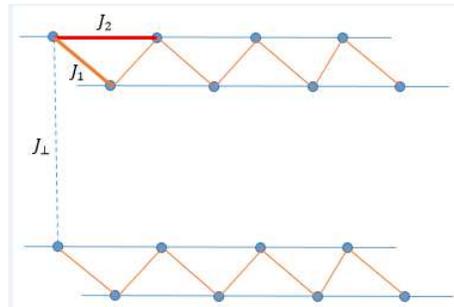}
 \caption{Schematic view of the interaction between the ions $V^{4+}$ in chain $\beta - \mathrm{T_eVO_4}$.}
 \label{f:2}
\end{figure}

%%%%%%%%%%%%%%%%%%%%%%%%%%%%%%%%%%%
Vanadium oxides with the $V^{4+}$ ions ($S=1/2$) are excellent model systems for one-dimensional 
spin-$1/2$ quantum magnets. Its structure consists of zig-zag chains, as show in Figure \ref{f:1}. 
A simplified scheme of this structure can be seen in Figure \ref{f:2}.

The interaction between nearest neighbors, represented by $J_1$, is weak in magnitude. Indeed, in Ref.\cite{ref17},
the authors have shown that in the temperature range of $5\,\mathrm{K}$ to $130\,\mathrm{K}$ the interaction is
dominated by $J_2$ and is of antiferromagnetic character, so $\vert J_1\vert \ll \vert J_2 \vert$. 
Therefore, we will take into account the interaction between next-nearest-neighbor represented for 
$J_2$ in the Figure \ref{f:2}. The interaction between the zig-zag chains $J_{\perp}$, is also weak 
and will be neglected here. At low temperatures ($T<5\,\mathrm{K}$) features three different magnetics 
characteristics. This is due to unknown origin of phase transition and our model $1-D$ is no longer 
very realistic. In short, in the temperature range of $5\,\mathrm{K}$ to $130\,\mathrm{K}$, we can consider this 
system as an infinite chain of spin-$1/2$ with antiferromagnetic interaction modeled by Heisenberg Hamiltonian (\ref{eq0}).

\begin{figure}
 \centering
 \includegraphics[width=7.5 cm]{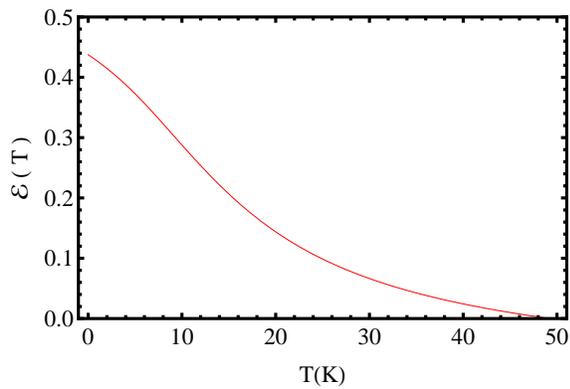}
 \caption{Thermal entanglement compound $\beta - \mathrm{T_eVO_4}$. 
 Mo\-de\-led as an infinite chain of spin-$1/2$.}
 \label{f:3}
\end{figure}

The value found in \cite{ref17} for the exchange interaction in this system is $J/k_B = -21.4\,\mathrm{K}$. 
Replacing it in (\ref{eq16}) we obtain the thermal entanglement system and we found the critical temperature 
of entanglement. The Figure \ref{f:3} shows the entanglement as a function of temperature for this compound.
This allows us to obtain materials with any critical temperature of 
entanglement previously desired. Thus, we are able to seek materials with appropriate 
$J$, as already was conjectured in \cite{ref10}. This is outstandingly interesting for the study of 
processing and transmission of quantum information.

It is a difficult task to determinate experimentally 
if a state is entangled or not. A widely used method for entanglement 
detection is the use of Entanglement Witnesses (EW's) \cite{ref18, ref19}. An 
observable W can be used as an EW if $\mathrm{Tr}(\rho W)<0$, when $\rho$ is an
entangled state. When $\mathrm{Tr}(\rho W)\geq0$, $\rho$ may or may not be entangled. 
Magnetic susceptibility was proposed as an EW \cite{ref6}, and several experimental
results were obtained within this framework \cite{ref20, ref21, ref22, ref9, ref23}.
At this point it easy to apply this method. Just do use the 
$EW=6k_BT\left(\frac{\overline{\chi}}{(g\mu_B)^2N} \right)-1$ \cite{ref6} and experimentally 
measured values of $\chi$ obtained in Ref. \cite{ref17}. Thus, according to the definition 
of Entanglement Witness, the inequality $6k_BT\left(\frac{\overline{\chi}}{(g\mu_B)^2N} \right)<1$
determines the existence of entanglement. However, $6k_BT\left(\frac{\overline{\chi}}{(g\mu_B)^2N} \right)\geq1$
does not assure separability. This method reveals a critical temperature of $T\simeq 31\mathrm{K}$.
This show that some entangled states were not detected by the EW, since our previous results show entanglement
in the system up to $T\simeq 47\mathrm{K}$ (see Figure \ref{f:3}).

Finally, this entanglement measure can be used to detect quantum phase transitions in Heisenberg chains 
with a external magnetic field and or anisotropy. In the present work we study a Heisenberg model without external 
magnetic field and anisotropy. Therefore do not exhibit quantum phase transition. This is a task of a later study.

%%%%%%%%%%%%%%%%%%%%%%%%%%%%%%%%%%%%%
\section{Conclusion}
\label{sec:Conc}
%%%%%%%%%%%%%%%%%%%%%%%%%%%%%%%%%%%%%
In conclusion, we have shown that an infinite chain entanglement
spins has a nonzero value in both $T=0$ and $T>0$. The critical
temperature of compound $\beta - \mathrm{T_eVO_4}$ was obtained.
Calculations were obtained analytically, using the distance 
between states and the Hilbert-Schmidt norm as a measure of entanglement. Our results allow 
us to conclude that the critical temperature of entanglement
increases directly with the increase in the exchange constant 
$J$. This allows us to conclude that the exchange constant 
can be seen as a ``shield'' against decoherence of entanglement, 
due to increased temperature.

%%%%%%%%%%%%%%%%%%%%%%%%%%%%%%%%%%%%%%%%%%%%%%%%%%%%%%%%%%
\section*{Acknowledgments}
The author would like to thank Profs. Daniel H.T. Franco and Oswaldo M. Del Cima for the many discussions 
and the fruitful collaborations. The author also would like to thank Prof. G\' eza T\'oth by your patient reading of the work and valuable suggestions. This work was partially supported by the Brazilian agency FAPEMIG.  
%%%%%%%%%%%%%%%%%%%%%%%%%%%%%%%%%%%%%%%%%%%%%%%%%%%%%%%%%%

%%%%%%%%%%%%%%%%%%%%%%%%%%%%%%%%%%%%%%%%%%%%%%%%%%%%%%%%%%

\bibliographystyle{apsrev4-1}
\bibliography{ensembles}

\end{document}